# Departure from Babinet principle in metasurfaces supported by subwavelength dielectric slabs

GIORGIO BIASIOL[1] AND SIMONE ZANOTTO[2*]

[1]*Istituto Officina dei Materiali – CNR, Laboratorio TASC, Basovizza (TS) (Italy)*
[2]*Istituto Nanoscienze – CNR, and Laboratorio NEST, Scuola Normale Superiore, Piazza San Silvestro 12, 56127 Pisa (Italy)*
*Corresponding author: simone.zanotto@nano.cnr.it

**Symmetry principles and theorems are of crucial importance in optics. Indeed, from one side they allow to get direct insights into phenomena by eliminating unphysical interpretations; from the other side, they guide the designer of photonic components by narrowing down the parameter space of design variables. In this Letter we illustrate a significant departure from the Babinet spectral complementarity in a very common and technologically relevant situation: that of a patterned conducting screen placed on a subwavelength dielectric slab. The symmetry property predicted by Babinet theorem is correctly recovered for pairs of geometrically complementary – but less realistic in terms of applications – free-standing patterned screens. Our analysis merges experimental data with fully vectorial electromagnetic modeling, and provides also an alternative form of Babinet theorem that highlights a connection with the concept of electromagnetic duality.**

The last decade has experienced a surge of interest in the science and technology of metasurfaces[1]. This large success is well deserved, since outstanding results about wave control have been attained by more and more elaborate mastering of the electromagnetic response of individual resonators, and of resonator arrays, that constitute metasurfaces. A non-exhaustive list includes lenses, holograms, polarization handling objects, emitters, filters, non-reciprocal components, and tunable elements[2–5]. However, the design of metasurfaces often involves quite elaborate processes, where parameter space exploration or – more recently – automated design techniques are often time-consuming processes[6,7]. To improve the metasurface synthesis procedures it is highly desirable to have fundamental rules, or symmetry constraints, that help identifying which are the parameters that are relevant for addressing a given target[8,9].

In this view, the Babinet principle is a formidable rule, since it links the response of a given impenetrable patterned screen to that of the complementary screen[10,11]. Since under certain circumstances metasurfaces are realized by means of thin metals operating below the plasma frequency, and are hence essentially impenetrable patterned screens, the Babinet principle is a powerful tool for the designer, as it doubles the yield of a given design process[12]. In general, it may be useful to identify more efficient devices based on light scattering and generation in structured media[13–17]. Recent works have also provided generalization of the Babinet principle to partially coherent light[18] and finite-conductivity interfaces[19], where it gives important guidelines towards the realization of frequency-independent screens, coherent perfect absorbers, polarization switches, integrated photonic components and terahertz reconfigurable filters[20–25]. Moreover, screens close to self-complementarity show interesting scaling properties and percolation effects[26].

Regarded as a principle when stated in terms of scalar field amplitudes, but being actually a rigorous theorem when dealing with the vectorial electromagnetic field, Babinet complementarity states that

$$\mathbf{E} - c\mathbf{B}_c = \mathbf{E}_0 \; ; \qquad c\mathbf{B} + \mathbf{E}_c = c\mathbf{B}_0 \qquad (1)$$

where $\mathbf{E}$, $\mathbf{E}_c$, and $\mathbf{E}_0$ are, respectively, the electric field scattered from the "original" screen, the electric field scattered from the complementary screen, and the electric field in absence of any screen. Subscripts have the same meaning also for the magnetic induction field $\mathbf{B}$; "c" identifies the speed of light. The relations in (1) are valid in the half-space $z > 0$, having assumed that the sources are in the half-space $z < 0$ and that the screen is placed at $z = 0$. Importantly, Eqs. (1), which we will name *Babinet theorem*, are valid only if the screen is perfectly conducting, if it has vanishing thickness, and if it is embedded in a symmetric infinitely-extended environment.

It is interesting to notice that Eqs. (1) have a form that can be enlightened by resorting to the electromagnetic duality transformations. Indeed, if one chooses the duality transformation in free space that operates as follows: $\mathbf{E} \rightarrow -c\mathbf{B}$, $c\mathbf{B} \rightarrow \mathbf{E}$, and if the electric and magnetic fields are joined together in a vector $\mathbf{F} = [\mathbf{E}, c\mathbf{B}]$, the Babinet theorem is written as

$$\mathbf{F} + \hat{\mathbf{F}}_c = \mathbf{F}_0 \qquad (2)$$

where $\hat{\mathbf{F}}$ denotes the dual of $\mathbf{F}$, and the $c$ and $0$ subscripts have the same meaning as above. This is a compact and handy form that bridges the concepts of geometric complementarity (which has been sometimes referred to as *Babinet duality*[27]) and electromagnetic duality. To our knowledge, this form of the Babinet theorem has never been reported so far.

Eqs. (1-2) can be manipulated to obtain an important corollary of Babinet theorem, which states that under certain conditions[1] the transmittance of a screen plus that of its complement equal unity:

$$T + T_c = 1. \qquad (3)$$

However, the power of this constraint is hindered by the strong hypotheses needed. From one side, the screen is required to be of vanishing thickness and to be perfectly conducting. While this requirement may be quite crude for visible metasurfaces[28], it however does not pose serious problems when moving towards longer wavelengths.

Rather, the aspect on which we are focusing our attention is that about the medium that embeds the metasurface, which must be the same in both $z < 0$ and $z > 0$ regions in order for Eqs. (1-3) to hold. Actually, a generalization for different embedding media and oblique incidence exists[29], but it requires that the embedding media are infinitely extended and homogeneous. Hence, such theory cannot be employed to the study of a relevant system for applications: that of a dielectric-slab-supported metallic metasurface. When the meta-atom array is placed on a finite-thickness, and possibly wavelength-comparable, dielectric slab, multiple reflections and/or guided wave effects can play a significant role, like in Salisbury screen geometries or guided-mode resonance filters[30]. Furthermore, membrane-supported metasurfaces are definitely of interest for optomechanically tunable devices[31]. In this article we report about the deviation from the spectral complementarity that would follow from Babinet theorem in a family of slab-supported complementary metasurfaces, which we have fabricated with gallium arsenide membrane technology, and measured in the mid-infrared range. In parallel we have developed an accurate numerical model, which we employed to extrapolate the behavior of the metasurface when the slab thickness is made to vanish. We observed that Babinet theorem is recovered only if the dielectric slab that supports the metasurface has a very strongly subwavelength thickness.

Figure 1a illustrates the metasurface under analysis. It consists of a gallium arsenide (GaAs) membrane, whose thickness $h$ in the experiment is close to 1.7 μm; that membrane supports an array of gold patches arranged in a square lattice with periodicity 4.15 μm. Gold thickness (approx. 50 nm) is much less than $h$. The fabrication process, described in details elsewhere[32], involves the epitaxial growth of GaAs and AlGaAs waveguide and sacrificial layers, electron beam lithography, metal deposition and lift-off, and wet etching procedures. The membrane is wide several tens of times the wavelength involved in the analysis. The gold patches are either disconnected, having a square shape, or are connected, leaving void squares, as illustrated in Fig. 1b. The shape is parametrized by means of a number $f \in [-1,1]$, which is linearly connected to the metal (or void) square side length. $f = -1$ means void squares of vanishing size; $f = 0$ means equally sized metal and void squares; $f = 1$ means metal patches of vanishing size.

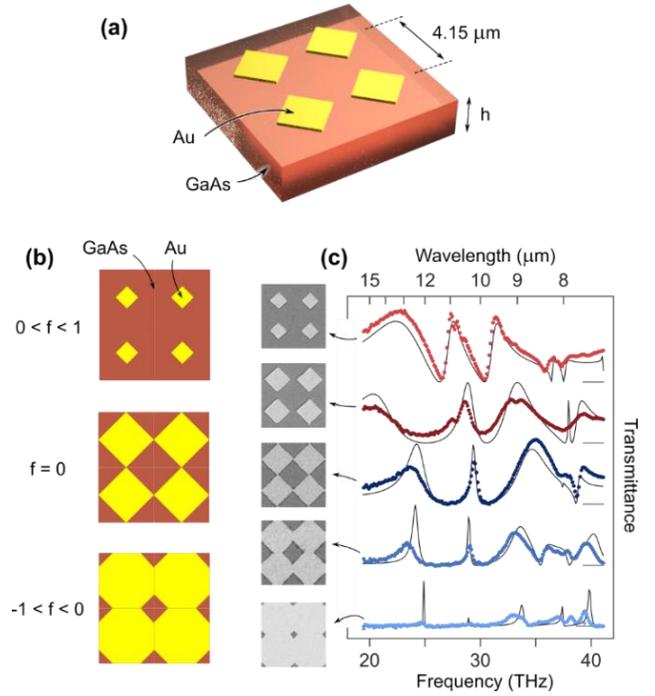

Fig. 1. Schematic of the complementary metasurface family and transmittance spectra. a) Details of the metasurface in a 3d rendering. b) Top-view of four unit cells with indication of the ranges of the parameter $f$ that identifies connected/disconnected metallic patterns. $f = 0$ is the self-complementary structure. c) Measured (colored dotted traces) and calculated (solid black lines) transmission spectra of five fabricated samples, with $f$ = -0.54, -0.29, -0.035, 0.23, 0.45 (from bottom to top). The corresponding scanning electron micrographs are also reported. The measured data are scaled and offset by arbitrary factors, while the calculated traces are absolute transmittance values and are offset by one unit.

We have fabricated five samples with $f$ = -0.54, -0.29, -0.035, 0.23, 0.45. Notice that the values are almost, but not exactly, symmetrical with respect to zero; this is due to lithography tolerances. In Fig. 1c we report the transmittance spectra collected from the samples by means of Fourier-transform infrared spectroscopy (FTIR); each spectrum is accompanied by a scanning electron microscope image of the corresponding sample. A magnified image of the $f$ = -0.035 sample is displayed as Visualization 1. The spectra are vertically offset for clarity; the experimental data are reported as dotted colored traces, while the result of a numerical model is reported as a solid black line.

If one applies to our series of metasurfaces the Babinet theorem corollary, Eq. (3), the spectra are expected to fulfill the relation $T(\nu; f) = 1 - T(\nu; -f)$, where $\nu$ is the radiation frequency and $f$ is the geometrical parameter. However, the data in Fig. 1c do not seem to comply with that relation. The observed mismatch could nonetheless be attributed, at first instance, to experimental issues: indeed, the values of $f$ are not exactly distributed in a symmetric way with respect to zero, and the thicknesses of the gallium arsenide membrane might have been different from sample to sample.

---

[1] Under the additional hypothesis that the screen has a pattern with a symmetry degree such that the scattering preserves polarization, and considering that the scattered E-fields are even with respect to $z$, one gets $t = -r_c$ and $t + t_c = 1$, where $t$ and $r$ are the transmission/reflection amplitude coefficients and $c$ labels the complementary screen. Adding the constraint of energy conservation (which holds if the screen operates in the diffractionless metasurface regime; dissipation being already ruled out by the hypothesis of perfect conduction) one gets the polarization-independent relation for the intensities $T + T_c = 1$.

For this reason we developed a numerical model to get further insight into the observed phenomena. We have solved the Maxwell equation by means of the finite-element modeling (FEM) software COMSOL, where a single unit cell with periodic boundary conditions is analyzed, and where the metal is treated as a perfectly conducting boundary. For numerical stability reasons, the conducting screen has a thickness of 100 nm. GaAs is modeled as a nondispersive dielectric with $\varepsilon_D = 10$. Source fields are injected into the simulation volume by means of port conditions; according to the experimental situation, plane waves at normal incidence polarized along the lattice periodicity direction are employed. After appropriate fitting of the membrane thickness (which ranges from 1.6 to 2 μm depending on the sample), the numerical data match well the experimental ones, apart from minor deviations in the peak amplitude and resonance lineshapes, which may be attributed to sample inhomogeneity (i.e. membrane wrinkles, thickness fluctuations of the membrane within a sample). This provided us a reliable simulation tool to accurately explore the electromagnetic response of the prototypical family of slab-supported, complementary metasurfaces described in Fig. 1a.

The first analysis concerned a series of metasurfaces with different values of $f$, distributed in a rigorously symmetrical way around zero ($f = \pm[\,0.03, 0.15, 0.3, 0.45, 0.63\,]$), for a fixed thickness of $t = 1.8$ μm. The transmittance spectra are reported in Fig. 2, where the curves are vertically offset for clarity. It can be immediately recognized that the relation $T(\nu; f) = 1 - T(\nu; -f)$ does not hold.

Even a "limit" equality for $f \to 0^\pm$ does not occur. Rather, it can be recognized that the curves are grouped in two continuously-evolving families: one for $f > 0$, another for $f < 0$. Within these two families, the metasurface resonances can be traced back to the guided-mode resonances, that are labeled in the Figure as TX$_j^{[m,n]}$, and that are well-defined for $|f| \to 1$. Indeed, under that limit, the metasurface becomes more and more similar to an unpatterned slab waveguide, either of the air/dielectric/air (ADA) type for $f \to 1$, or of the metal/dielectric/air (MDA) type for $f \to -1$. These waveguides support TE and TM modes of order $j$ that can be excited if the Bragg condition $|m\mathbf{g}_1 + n\mathbf{g}_2| = \beta$ is satisfied. Here, $m$ and $n$ are relative integers, $\mathbf{g}_1 = (2\pi/a, 0)$ and $\mathbf{g}_2 = (0, 2\pi/a)$ are the reciprocal space unit vectors (where $a$ is the pattern periodicity), and $\beta = 2\pi n_{\text{eff}}/\lambda_0$ is the guided mode wavevector. In Figure 2 we report the lowest-order guided mode resonances for both ADA and MDA waveguides: they show very strong correlation with the spectral features observed in the transmission spectra of the metasurfaces with $|f| = 0.63$. Larger values of $|f|$ are expected to lead to even stronger matching; however, they are quite difficult to simulate with the FEM because of the presence of strongly subwavelength metallic holes or patches. Notice the clear Fano lineshapes occurring for $f = 0.63$, where the guided mode resonances stand out of a Fabry-Pérot background dictated by the dielectric slab parallel-interface scattering. The experimental data of Fig. 1, further supported by the numerical analysis of Fig. 2, clearly show that the Babinet theorem, in its form of Eq. (3), does not hold for complementary metasurfaces backed by a finite-thickness dielectric material.

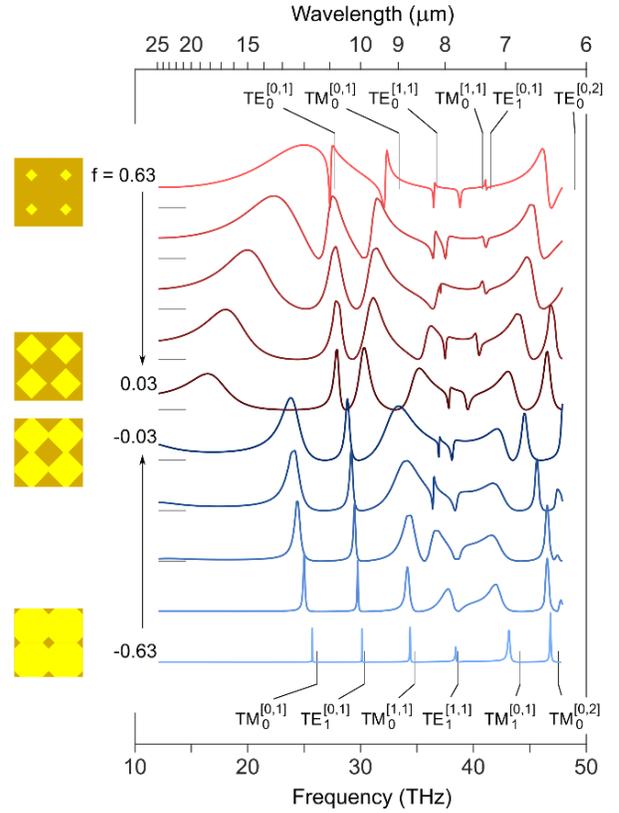

Fig. 2. Violation of Babinet theorem observed in a series of complementary slab-supported metasurfaces. The transmittance spectra (absolute transmittance, each offset by 0.8) do not comply with the Babinet theorem requirement $T(\nu; f) = 1 - T(\nu; -f)$, where $\nu$ is the frequency and $f$ is the complementarity-controlling geometrical parameter. The labels TE$_j^{[m,n]}$ and TM$_j^{[m,n]}$ identify the guided mode resonances originating from the slab waveguide modes. A detailed analysis is provided in the main text.

The Babinet principle violation process retains clear fingerprints of the waveguiding nature of the dielectric slab that supports the metallic pattern. These observations (i.e. spectral complementarity violation and guided mode resonance fingerprints) do not stem from the absence of a mirror symmetry plane. Additional simulations (not shown) revealed that they remain valid also if a finite-thickness dielectric slab is symmetrically embedding the patterned conducting screen.

With the numerical tool at our disposal, we can however explore what happens if the hypotheses of the Babinet theorem are recovered: namely, we considered a pair of metallic complementary patterns, and progressively reduced the thickness of the supporting dielectric slab. The results are plotted in Figure 3, where the red traces (upper half) are calculated for a "disconnected" metasurface ($f = 0.25$) while the blue traces (lower half) are calculated for the corresponding "connected" metasurface ($f = -0.25$). In the upper half of the plot we also report the transmittance complement, $1 - T$, of the traces of the lower half, to highlight what is the entity of the violation of the relation $T(f) = 1 - T(-f)$. Noticeably, the relation is visibly violated also for thicknesses as small as 0.2 μm. At this thickness value, the violation concerns mostly the lineshape (of the resonance occurring at ≈ 7.5 μm), while for larger thicknesses the violation concerns also the resonance position (according to the observations of Fig. 2). Instead, the transmittance curves for $h = 0$ satisfy to a much better extent the relation $T(f) = 1 - T(-f)$. Here, the small discrepancy can be attributed to the finite value of the metal thickness (0.1 μm, see above for the reason of this choice). Larger, but still subwavelength, metal thicknesses also lead to departure from Babinet complementarity, in the form of peak narrowing (broadening) for the cases $f = -0.25$ ($f = 0.25$), respectively (not shown).

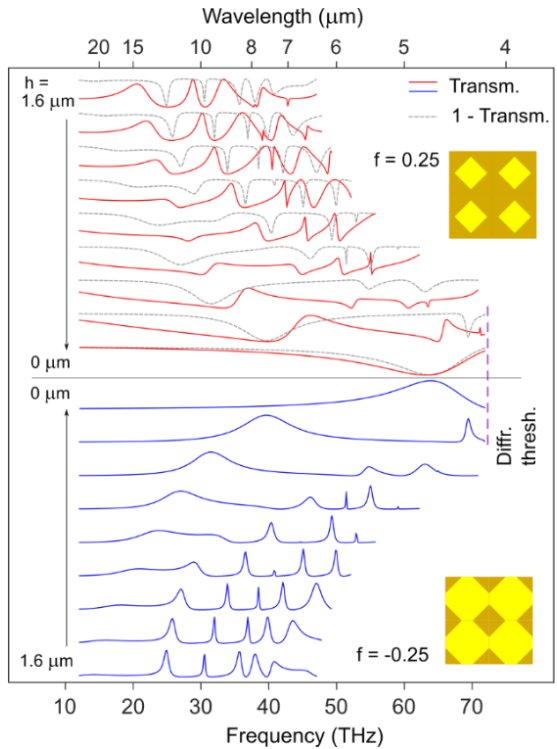

Fig. 3. Recovery of Babinet theorem in a series of complementary metasurfaces supported by dielectric slabs of different thicknesses $h$. ($h$ = 1.6, 1.4, 1.2, 1, 0.8, 0.6, 0.4, 0.2, 0 μm). As the slab thickness approaches zero, complementary transmission spectra are observed. The spectra (absolute transmittance) are offset by 1.2 units.

In summary, we have illustrated theoretically and experimentally a significant departure from the expectations of Babinet principle when a dielectric slab, even of subwavelength thickness, is placed in close vicinity to complementary patterned conducting screens. Stronger deviations are observed in the vicinity of spectral structures characteristic of guided mode resonances. In stating the Babinet principle (or better, Babinet theorem) we have also noticed that it can be expressed in a form that draws a direct connection between geometric complementarity and electromagnetic duality. We believe that the results presented in this Letter may be of help for the design of frequency-selective filters, metasurface membranes and in general for complex wave-handling devices, where the designer can be strongly supported by a correct identification of the presence or absence of symmetries induced by electromagnetic duality and by geometric complementarity.

**Acknowledgment**

The authors wish to thank Dr. Riccardo Degl'Innocenti (Lancaster University) who contributed at an early stage of the project. A critical reading of the manuscript from Prof. Alessandro Tredicucci (Università di Pisa) and Prof. Giuseppe C. La Rocca (Scuola Normale Superiore, Pisa) is very kindly acknowledged.

**Disclosures**

The authors declare no conflicts of interest.